\documentclass{article}
\usepackage{spconf,amsmath,graphicx,hyperref}

\usepackage{booktabs,multirow}

\usepackage{xcolor}

\usepackage{pgfplots}
\usepgfplotslibrary{groupplots} 
\pgfplotsset{compat=1.18}
\usepgfplotslibrary{groupplots}
\pgfplotsset{compat=newest}
\usepackage{newtxtext}
\usepackage{newtxmath}

\definecolor{myblue}{RGB}{65, 105, 225} 
\definecolor{mygray}{RGB}{200, 200, 200} 
\title{DTT-BSR: GAN-Based DTTNet with RoPE Transformer Enhancement for Music Source Restoration}

\name{\parbox{\linewidth}{\centering
    Shihong Tan$^{1,\star}$, Haoyu Wang$^{2,4,\star}$, Youran Ni$^{1}$, Yingzhao Hou$^{1}$, Jiayue Luo$^{1}$,\\
      \textit{Zipei Hu$^{1}$, Han Dou$^{1}$, Zerui Han$^{4}$, Ningning Pan$^{2}$, Yuzhu Wang$^{3}$, Gongping Huang$^{1,\dagger}$}}
      \thanks{$^{\star}$These authors contributed equally to this work.}
      \thanks{$^{\dagger}$Corresponding author.}
      \thanks{This work was supported by the National Natural Science Foundation of China under Grant 62471340 and 62401479 and Sichuan Science and Technology Program under Grant 2026NFSFC1430.}
      }
\address{$^{1}$School of Electronic Information, Wuhan University, China \\
         $^{2}$Southwestern University of Finance and Economics, China\\
         $^{3}$Tampere University, Finland\\
         $^{4}$MiLM Plus, Xiaomi Inc., China}

\begin{document}
%
\maketitle

\begin{abstract}
Music source restoration (MSR) aims to recover unprocessed stems from mixed and mastered recordings. The challenge lies in both separating overlapping sources and reconstructing signals degraded by production effects such as compression and reverberation.
We therefore propose DTT-BSR, a hybrid generative adversarial network (GAN) combining rotary positional embeddings (RoPE) transformer for long-term temporal modeling with dual-path band-split recurrent neural network (RNN) for multi-resolution spectral processing. 
Our model achieved 3rd place on the objective leaderboard and 4th place on the subjective leaderboard on the ICASSP 2026 MSR Challenge, demonstrating exceptional generation fidelity and semantic alignment with a compact size of $7.1$M parameters.
\end{abstract}
\begin{keywords}
Music source restoration, GAN, Music source separation
\end{keywords}
\section{Introduction}
\label{sec:intro}

Music source restoration (MSR)~\cite{zangMusicSourceRestoration2025} extends music source separation (MSS) by requiring both source isolation and restoration of signals degraded during music production. 
Inspired by Dual-Path TFC-TDF UNet (DTTNet)~\cite{chenMusicSourceSeparation2024} and Band-Split RoPE Transformer (BSRoFormer)~\cite{luMusicSourceSeparation2023},
We propose a generative adversarial network (GAN) based adaptation of the DTTNet with band-sequence modeling and rotary positional embeddings (RoPE)~\cite{su2024roformer} transformer bottleneck (DTT-BSR).
Specifically, DTTNet is adopted as backbone for its efficient U-Net structure, with RoPE transformer blocks equipped to capture long-term dependencies and dual-path recurrent neural network (RNN) modules for fine-grained spectral features. 
This approach bridges discriminative separation and generative restoration within a single, end-to-end framework.

Our approach achieves competitive performance in the ICASSP 2026 MSR Challenge, ranking 3rd on objective metrics and 4th on subjective evaluation. Code\footnote{https://github.com/OrigamiShido/DTT-BSR} and pretrained weights\footnote{https://huggingface.co/OrigamiShido/MSRChallenge-ACDC} are publicly available.

\begin{table*}[t!]
    \centering
    \renewcommand{\arraystretch}{0.86}
    \caption{Overall Results on the Official Test Set}
    \label{table:result-test} 
    \vspace{-0.0cm}
    \setlength{\tabcolsep}{3.5pt} 
    \begin{tabular}{lcccccc} 
        \toprule
    Method & MMSNR & Zimtohrli & FAD-CLAP & MOS\_Sep & MOS\_Rankrestoration & MOS\_overall \\
        \midrule
        DTT-BSR & 1.4520 & 0.0182 & 0.2907 & 3.5425 & 2.4768 & 2.5412 \\
        \bottomrule
    \end{tabular}
    \vspace{-0.6em}
    \caption{Results per Stem on the Official Test Set}
    \label{table:result-stem} 
    \vspace{-0.0cm}
    \setlength{\tabcolsep}{3.5pt}
    \begin{tabular}{cccccccccc}
        \toprule
        Method & Metrics & Vocals & Gtr. & Key. & Synth & Bass & Drums & Perc. & Orch. \\
        \midrule
        \multirow{3}{*}{DTT-BSR} & MMSNR     & 1.0494 & 1.1410 & 1.8237 & 0.9508 & 2.8572 & 2.7304 & 0.0171 & 1.0464 \\
                                 & Zimtohrli & 0.0195 & 0.0165 & 0.0167 & 0.0183 & 0.0142 & 0.0185 & 0.0230 & 0.0189 \\
                                 & FAD-CLAP  & 0.4186 & 0.3386 & 0.4257 & 0.6060 & 0.6026 & 0.4039 & 0.9606 & 0.3275 \\
        \bottomrule
    \end{tabular}
    \vspace{-1.em} 
\end{table*}

\section{methodology}
\label{sec:method}
\subsection{Model Architecture}

\begin{figure}[htb]
  \centering
  \centerline{\includegraphics[width=8.5cm]{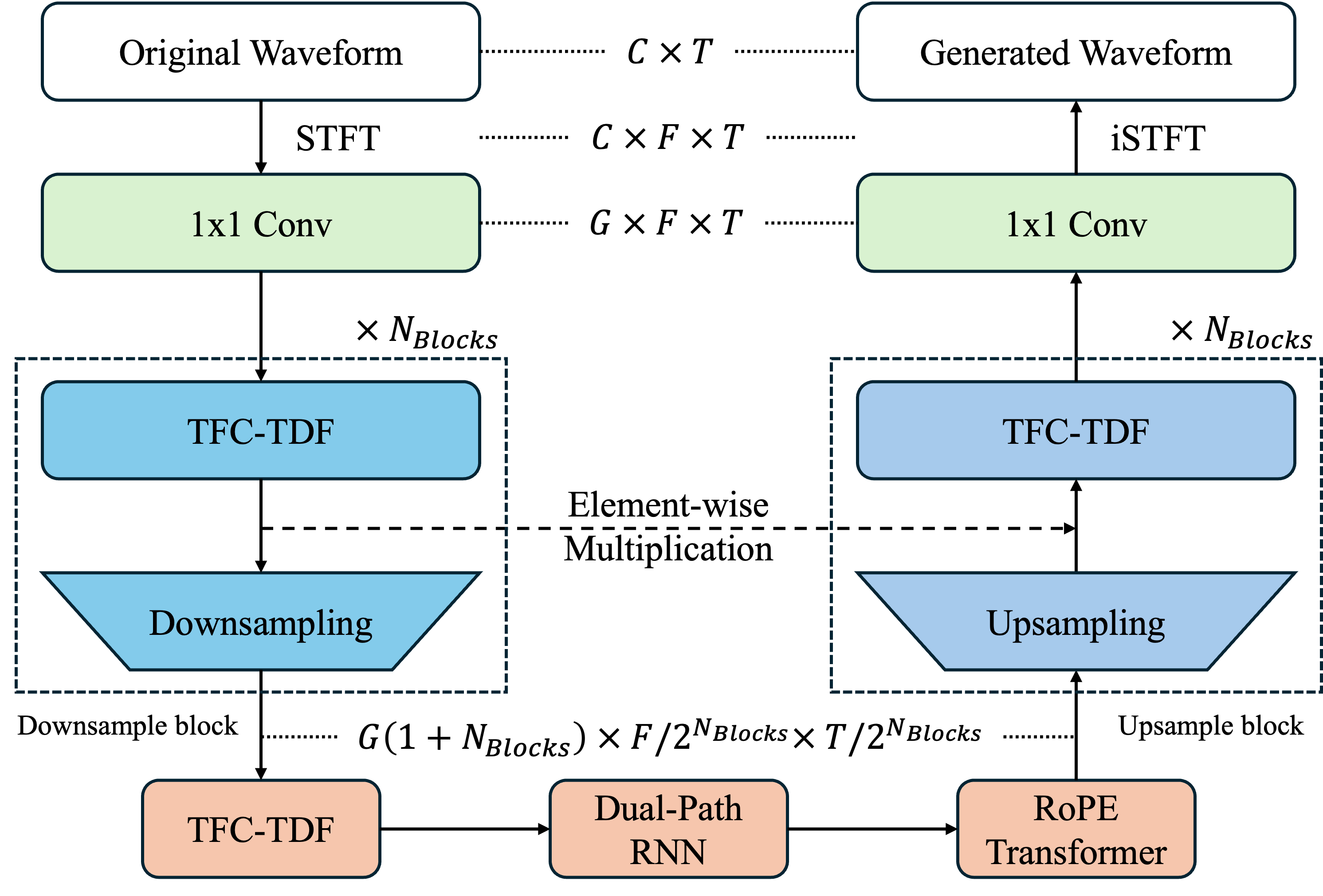}}
\caption{Our Proposed Model Architecture}
\label{fig:arc}
\end{figure}

Figure \ref{fig:arc} shows the framework of the proposed method, DTT-BSR, where DTTNet is adopted as the backbone, with RoPE transformer blocks incorporated to capture long-term dependence, and dual-path RNN block for fine-grained time-frequency feature extraction.
The $C$-channel time-domain waveform $\textbf{W}^{C\times T}$ is first processed through short-time Fourier transformation (STFT), which yields $\textbf{S}^{C\times T\times F}$. 
Then, a $1 \times 1$ convolution layer is applied, which yields a $G$-dimensional feature, 
which is then processed through $N_{blocks}$ of downsample blocks, each containing a TimeFrequency Convolutions Time-Distributed Fully-connected(TFC-TDF) block~\cite{kim2023sound} and a convolution layer that halves the feature map and increase the feature dimensions by $G$.
The high-dimensional features are subsequently processed by another TFC-TDF block, a dual-path RNN block\cite{chenMusicSourceSeparation2024}, and a RoPE transformer block.
The bottle-neck latent features then go through $N_{Blocks}$ of upsample blocks, which contains a convolutional layer and a TFC-TDF block.
We apply skip connection mechanism to improve the connectivity, where element-wise multiplication is performed after each upsampling block.
Finally, a $1 \times 1$ convolution layer is added to project the feature dimension $G$ back to $C$.
An inverse STFT operation is conducted to transform the spectrum to the generated waveform.


The proposed model is trained using a composite objective that combines regression losses with adversarial feedback from the multi-frequency discriminator of EnCodec~\cite{défossez2022highfidelityneuralaudio}. Specifically, the joint loss function $\mathcal{L}$ comprises:
The Multi-Mel STFT loss, $\mathcal{L}_\mathrm{MMS}$, computes L1 distance between magnitude spectrograms at multiple window sizes;
the adversarial loss $\mathcal{L}_{\text{adv}}$ uses a hinge loss formulation to improve perceptual quality by encouraging the generator to produce realistic samples that fool the discriminator;
the feature matching loss $\mathcal{L}_\mathrm{feat}$ measures L1 distance between discriminator feature maps of real and generated samples.
Hence, $\mathcal{L}$ can be calculated by

\begin{equation}
\mathcal{L}=\lambda_\mathrm{MMS}\mathcal{L}_\mathrm{MMS}+\lambda_\mathrm{adv}\mathcal{L}_\mathrm{adv}+\lambda_\mathrm{feat}\mathcal{L}_\mathrm{feat} 
\label{eq:loss}
\end{equation}
where $\lambda_\mathrm{MMS}$, $\lambda_\mathrm{adv}$, and $\lambda_\mathrm{feat}$ are weights that balance the contribution of each loss term. In our experiments, these hyperparameters were set to $45.0$, $2.0$, and $4.0$, respectively.

\section{experiments}
\label{sec:exp}

\subsection{Model Training and Evaluation}

We only use RawStems\cite{zangMusicSourceRestoration2025} as the training set, which includes $578$ songs containing all 8 target stems with a total $354.13$ hours of length.
Dynamic Range Compression (using compressor and limiter), Harmonic Distortion, Reverb, and Random Audio Resample are applied as data augmentation method.
Hanning window is used for conducting STFT, with the window length of $2048$, hop length of $512$.
We set $N_{Blocks}$ at $2$, and the convolution kernel size of TFC-TDF module at $(3,3)$.
The number of layers of the dual-path module is set to be $4$ and number of heads set to be $2$. 
The RoPE parameters are set to repeat $2$ times with $8$ heads, $2$ time and frequency transformer modules each, with $0.1$ of dropout rate.
Final generator parameter amount is $7.1$M.
We use AdamW for optimizer and set the initial learning rate of $0.002$.
A single NVIDIA RTX 5090 is used to train each stem with a batch size of $2$, for a total of $1$ million steps each stem, and the training takes about $26$ hours.

We use the officially released MSRBench~\cite{zang2025msrbench} for model evaluation. For checkpoint selection, we evaluate all saved checkpoints on the validation set using Multi-Mel SNR (MMSNR), FAD-CLAP\cite{wu2024largescalecontrastivelanguageaudiopretraining}, and Zimtohrli\cite{alakuijalaZimtohrliEfficientPsychoacoustic2025}. The best-performing checkpoint is then evaluated on the official test set.

\subsection{Results and Analysis}

    
    

\begin{table}[t!]
    \centering
    \renewcommand{\arraystretch}{0.85}
    \vspace{-0.2cm}
    \caption{Overall Objective Results on MSRBench}
    \label{table:result-validation}
    \vspace{-0cm}
    \setlength{\tabcolsep}{3.5pt}
    \begin{tabular}{cccc}
    \toprule
    Method & MMSNR ($\uparrow$) & Zimtohrli ($\downarrow$) & FAD-CLAP ($\downarrow$) \\
    \midrule
    DTT-BSR         & \textbf{0.5011}             & 0.0216                            & \textbf{0.5660}                  \\
    Baseline        & 0.4020                      & 0.0216                            & 0.7545                           \\
    \bottomrule
    \end{tabular}
    \vspace{-2.0em}
\end{table}

The model is tested with the official released test set.
Table \ref{table:result-test} shows the overall objective and subjective results and Table \ref{table:result-stem} details the per-stem results\cite{zang2026summaryinauguralmusicsource}. 
Throughout the $8$ target stems, our model performs better at Guitar, Keyboard and Orchestra stem, and outperforms by Zimtohrli, FAD-CLAP, and subjective evaluation, indicating that the proposed model improves perceptual quality and semantic alignment ability. According to our tested result on the validation set as in Table \ref{table:result-validation}, our model has a 24.62\% increase on MMSNR and 24.93\% decrease on FAD-CLAP, showing a significant improvement. 
DTT-BSR performs better at Medium-frequency stringed instruments while lacking notable performance on the other stems.
Hence, our model provides a promising solution on non-vocal instrument separations.

\section{conclusion}
\label{sec:conc}

We propose DTT-BSR, a DTTNet-based GAN generator which shows improvement in Music Source Restoration. With a high-efficiency DTTNet backbone integrated with RoPE transformer block, our proposed model achieves 3rd in Object metrics and 4th in Subject metrics in the ICASSP Music Source Restoration Challenge, highlighting the perceptual and semantic alignment ability, and non-vocal instrument performance.


\vfill\pagebreak

\bibliographystyle{ieeetr}
\bibliography{strings,refs}

@article{zang2025msrbench,
  title={MSRBench: A Benchmarking Dataset for Music Source Restoration},
  author={Zang, Yongyi and Hai, Jiarui and Ge, Wanying and Kong, Qiuqiang and Dai, Zheqi and Wang, Helin and Mitsufuji, Yuki and Plumbley, Mark D},
  journal={arXiv preprint arXiv:2510.10995},
  year={2025}
}

@INPROCEEDINGS{chenMusicSourceSeparation2024,
  author={Chen, Junyu and Vekkot, Susmitha and Shukla, Pancham},
  booktitle={ICASSP 2024 - 2024 IEEE International Conference on Acoustics, Speech and Signal Processing (ICASSP)}, 
  title={Music Source Separation Based on a Lightweight Deep Learning Framework (DTTNET: DUAL-PATH TFC-TDF UNET)}, 
  year={2024},
  volume={},
  number={},
  pages={656-660},
  doi={10.1109/ICASSP48485.2024.10448020}}

@article{kim2023sound,
  title={Sound Demixing Challenge 2023 Music Demixing Track Technical Report: TFC-TDF-UNet v3},
  author={Kim, Minseok and Lee, Jun Hyung and Jung, Soonyoung},
  journal={arXiv preprint arXiv:2306.09382},
  year={2023}
}

@misc{zangMusicSourceRestoration2025,
      title={Music Source Restoration}, 
      author={Yongyi Zang and Zheqi Dai and Mark D. Plumbley and Qiuqiang Kong},
      year={2025},
      eprint={2505.21827},
      archivePrefix={arXiv},
      primaryClass={cs.SD},
      url={https://arxiv.org/abs/2505.21827}, 
}

@INPROCEEDINGS{luMusicSourceSeparation2023,
  author={Lu, Wei-Tsung and Wang, Ju-Chiang and Kong, Qiuqiang and Hung, Yun-Ning},
  booktitle={ICASSP 2024 - 2024 IEEE International Conference on Acoustics, Speech and Signal Processing (ICASSP)}, 
  title={Music Source Separation With Band-Split Rope Transformer}, 
  year={2024},
  volume={},
  number={},
  pages={481-485},
  doi={10.1109/ICASSP48485.2024.10446843}}

@misc{alakuijalaZimtohrliEfficientPsychoacoustic2025,
      title={Zimtohrli: An Efficient Psychoacoustic Audio Similarity Metric}, 
      author={Jyrki Alakuijala and Martin Bruse and Sami Boukortt and Jozef Marus Coldenhoff and Milos Cernak},
      year={2025},
      eprint={2509.26133},
      archivePrefix={arXiv},
      primaryClass={eess.AS},
      url={https://arxiv.org/abs/2509.26133}, 
}

@misc{wu2024largescalecontrastivelanguageaudiopretraining,
      title={Large-scale Contrastive Language-Audio Pretraining with Feature Fusion and Keyword-to-Caption Augmentation}, 
      author={Yusong Wu and Ke Chen and Tianyu Zhang and Yuchen Hui and Marianna Nezhurina and Taylor Berg-Kirkpatrick and Shlomo Dubnov},
      year={2024},
      eprint={2211.06687},
      archivePrefix={arXiv},
      primaryClass={cs.SD},
      url={https://arxiv.org/abs/2211.06687}, 
}

@article{su2024roformer,
    title={Roformer: Enhanced transformer with rotary position embedding},
    author={Su, Jianlin and Ahmed, Murtadha and Lu, Yu and Pan, Shengfeng and Bo, Wen and Liu, Yunfeng},
    journal={Neurocomputing},
    volume={568},
    pages={127063},
    year={2024},
    publisher={Elsevier}
}

@article{défossez2022highfidelityneuralaudio,
title={High Fidelity Neural Audio Compression},
author={Alexandre D{\'e}fossez and Jade Copet and Gabriel Synnaeve and Yossi Adi},
journal={Transactions on Machine Learning Research},
issn={2835-8856},
year={2023},
url={https://openreview.net/forum?id=ivCd8z8zR2},
}

@misc{zang2026summaryinauguralmusicsource,
      title={Summary of The Inaugural Music Source Restoration Challenge}, 
      author={Yongyi Zang and Jiarui Hai and Wanying Ge and Qiuqiang Kong and Zheqi Dai and Helin Wang and Yuki Mitsufuji and Mark D. Plumbley},
      year={2026},
      eprint={2601.04343},
      archivePrefix={arXiv},
      primaryClass={cs.SD},
      url={https://arxiv.org/abs/2601.04343}, 
}

\end{document}